\documentclass{ws-mpla}
\usepackage[super]{cite}
\usepackage{graphicx}
\usepackage{amsmath,amsfonts,euscript,amssymb}
\usepackage{epsfig}

\newcommand{\be}{\begin{equation}}
\newcommand{\ee}{\end{equation}}
\newcommand{\bea}{\begin{eqnarray}}
\newcommand{\eea}{\end{eqnarray}}

\begin{document}

\markboth{A.B. Arbuzov and A.E. Pavlov}
{Static Casimir Condensate of Conformal Scalar Field in Friedmann Universe}

\catchline{}{}{}{}{}

\title{Static Casimir Condensate of Conformal Scalar Field in Friedmann Universe}

\author{Andrej B. Arbuzov$^{1,2}$}

\address{$^1$Bogoliubov Laboratory of Theoretical Physics,
Joint Institute for Nuclear Research, \\ Joliot-Curie str. 6,
Dubna, 141980, Russia \\
$^2$Department of Higher Mathematics,
Dubna State University, \\ Universitetskaya str. 19,
Dubna, 141982, Russia \\
arbuzov@theor.jinr.ru}

\author{Alexander E. Pavlov$^3$}

\address{$^3$Institute of Mechanics and Energetics,
Russian State Agrarian University ---
Moscow Timiryazev Agricultural Academy, Moscow, 127550, Russia \\
alexpavlov60@mail.ru}

\maketitle


\begin{abstract}
The quantum Casimir condensate of a conformal massive scalar field
in a compact Friedmann universe is considered in the static approximation.
The Abel--Plana formula is used for renormalization of divergent series
in the condensate calculation. 
A differential relation between the static Casimir energy density
and static Casimir condensate is derived.

\end{abstract}

\ccode{PACS Nos.: 11.25.Hf, 11.30.Qc}

\section{Introduction}

Quantum field condensates are known to appear in systems with spontaneous
symmetry breaking such as the Standard Model and QCD. In particular,
the Coleman--Weinberg mechanism~\cite{Coleman:1973jx} allows emergence
of a finite field condensate starting from a conformal-invariant semi-classical
Lagrangian. The quantum condensate mechanism of the origin of Higgs boson
mass without explicit conformal symmetry breaking was proposed in
Ref.~\cite{Arbuzov:2016xte}.
But to define of the conformal field condensate value, an additional
physical condition is required. One of the possibilities is to relate
the emergent finite value of the condensate to certain boundary conditions
of the system like in the Casimir effect. In the present paper we will
consider the case of a scalar field in a closed manifold.
The topological Casimir effect will be treated.
A non-trivial topology of the space provides the rule
how to fix the renormalization condition in the calculations.
Contrary to the case of the standard Casimir effect, no border
or boundary condition is imposed. 

The Casimir effect plays an essential role in cosmological models and in
particle physics~\cite{MTrunov}. Besides of the Casimir energy,
the Casimir condensates of quantum fields deserve a study, especially
in the early stage of the Universe.
The Casimir condensate can be defined as vacuum mean value of two field
operators in one point of the manifold. It corresponds to a tadpole 
(vacuum bubble) Feynman diagram without external legs. This object
contributes to the universe vacuum energy and its treatment is
crucial for cosmological applications~\cite{Martin:2012bt}. 

In the conformally static universe the calculation of the energy density, pressure,
and the density number of the boson pairs created during the evolution of
the universe from the singular initial state was performed in 
Ref.~\cite{Mamaev:1976zb}.
However, the Casimir energy-momentum tensor of a boson field has singularity
at the zero scale factor value, $a=0$, which corresponds to the initial moment
of the universe evolution. To avoid the initial cosmological
singularity, one can exploit conformal transformations to the static universe
that is a sphere of the present day radius $a_0$. The conformal transformation
should be applied also to quantum fields simultaneously. The calculation of
the Casimir condensate should be considered in conformal variables also.

From the conformal cosmological conception, the universe is not expanded
but the elementary particles masses turn out to be depend on time,
see e.g. Ref.~\cite{Narlikar}.
The redshift of galactic spectrum and the modern Hubble diagram can be
interpreted~\cite{Zakharov:2010nf,PavlovMIPh,Pervushin:2017zfj}
within the General Relativity in conformal variables without introducing Dark Energy.
The transition to the conformal variables has a considerable advantage since
the initial cosmological singularity doesn't appear there.
The procedures of Hamiltonian reduction and deparametrization in Geometrodynamics
by passing to the conformal spatial metric, were recently analyzed
in Ref.~\cite{Arbuzov:2017col}.

There are various effective methods of regularization and renormalization
of ultraviolet divergences. In particular, several such methods were
developed to treat the Casimir effect in different cases, see e.g.
book~\cite{MTrunov}.
In particular for the case of the topological Casimir effect, extraction
of the finite value of a boson condensate can be achieved
by subtraction from a divergent sum defined in the Friedmann space
the corresponding divergent integral defined in the Minkowskian tangent space.
The Abel--Plana formula from the theory of analytical functions can be applied for
this purpose, see Ref.~\cite{Mamaev:1976zb}.

\section{Conformal massive scalar field}

The conformal spacetime metric of a Friedmann universe
${\cal M}=\mathbb{R}^1\times S^3$ reads
\be \label{ds}
ds^2=a^2(\eta)\left(-d\eta^2+d\chi^2+\sin^2\chi(d\theta^2+\sin^2\theta d\phi^2)\right),
\ee
where $\eta$ is the conformal time; $a(\eta)$ is the conformal scale factor
having the sense of the $S^3$-sphere radius; $\chi$, $\theta$, and $\phi$ are
dimensionless angular coordinates on the sphere.
The Lagrangian density of a massive scalar field $\varphi (x)$ with a conformal
coupling has the form~\cite{PChT,Penrose}
\be
{\cal L}=-\frac{1}{2}\sqrt{-g}(x)\left(g^{\mu\nu}(x)\nabla_\mu\varphi (x)
\nabla_\nu \varphi (x)+m^2\varphi^2(x)+\frac{1}{6}R(x)\varphi^2(x)\right).
\ee
Here, $g$ is the determinant of the spacetime metric; $g^{\mu\nu}$ are components
of the inverse metric tensor; $\nabla_\mu$ is the covariant derivative;
$m$ is the bare mass of the field.
The scalar curvature of the spacetime~(\ref{ds}) is
\be
R(\eta)=\frac{6}{a^3}(a''+a),
\ee
where the prime denotes differentiation with respect to $\eta$.
The Klein--Fock--Gordon equation
\be \label{fieldeq}
\left(-g^{\mu\nu}\nabla_\mu\nabla_\nu+m^2+\frac{1}{6}R\right)\varphi (x)=0
\ee
takes the form
\be \label{KG}
\varphi''+2\frac{a'}{a}\varphi'-\triangle\varphi
+ \left(m^2 a^2+\frac{a''}{a}+1\right)\varphi=0,
\ee
where $\triangle$ is the angular part of the Laplace operator on a 3-sphere with
the unit radius.

Let us perform the transformation to the conformal metric $\tilde{g}_{\mu\nu}(x)$
and the conformal field $\tilde\varphi (x)$ according to their conformal weights:
\be \label{trans}
g_{\mu\nu}(x)=\left(\frac{a(\eta)}{a_0}\right)^2\tilde{g}_{\mu\nu}(x),\qquad
\varphi (x)=\left(\frac{a_0}{a(\eta)}\right)\tilde\varphi (x).
\ee
The static spacetime interval corresponds to the present day radius $a_0$ of
the universe
\be\label{tildeds}
d\tilde{s}^2=a_0^2(-d\eta^2+d\chi^2+\sin^2\chi(d\theta^2+\sin^2\theta d\phi^2))
\ee
with the curvature $\tilde{R}=6/a_0^2$. Transformation~(\ref{trans})
leads to the Klein--Fock--Gordon equation for the conformal scalar field
$\tilde\varphi (x)$  with a variable mass $m(\eta)=m\cdot a(\eta)$
\be \label{varphitilde}
\tilde\varphi''-\triangle\tilde\varphi+(m^2a^2+1)\tilde\varphi=0.
\ee
The transition to the conformal variables leads to observable quantities with
regular behavior at $a=0$. In the corresponding quantum theory it yields finite
physical results~\cite{Mamaev:1976zb}.
The eigenfunctions of~(\ref{varphitilde}) can be presented in the factorized form
\be\label{bfx}
\tilde\varphi_J (x)=g_\lambda (\eta)Y_J ({\bf x}).
\ee
Here, the eigenfunctions $Y_J (\chi, \theta, \phi)$ of the Laplace--Beltrami
operator~\cite{MTrunov} are spherical functions
\be\label{eigen}
(\triangle+k^2_J)Y_J=0,
\ee
forming the orthonormal basis of unitary representations of the isometry group
of the $S^3$-sphere
\be
Y_J(\chi, \theta, \phi)=\frac{1}{\sqrt{\sin\chi}}
\sqrt{\frac{\lambda (\lambda+l)!}{(\lambda-l+1)!}}
P^{-l-1/2}_{\lambda-1/2}(\cos\chi)Y_{lM}(\theta,\phi),
\ee
the combined index $J\equiv \{\lambda, l, M\}$ runs over values
\be
\lambda=1,2,3,\ldots;\qquad l=0,1,2,\ldots,\lambda-1;\qquad -l\le M\le l.
\ee
The eigenvalues of Eq.~(\ref{eigen}) are $k_J^2=\lambda^2-1$.
The equation of the oscillator type for $g_\lambda (\eta)$ is obtained after
substitution of~(\ref{bfx}) into the Klein--Fock--Gordon equation~(\ref{varphitilde})
\be\label{glambda}
\frac{d^2}{d\eta^2}g_\lambda (\eta)+\omega^2(\eta)g_\lambda (\eta)=0,\qquad
\omega^2 (\eta)\equiv\lambda^2+m^2a^2(\eta).
\ee
To find solutions of the above equation, it is necessary to choose a definite
evolution law for the scale factor $a(\eta)$~\cite{Mamaev:1976zb}.

The functions
\be
\tilde\varphi_J^{(+)}(x)=\frac{1}{\sqrt{2}}g_\lambda(\eta)Y_J^{*}({\bf x}),\qquad
\tilde\varphi_J^{(-)}(x)=\frac{1}{\sqrt{2}}g_\lambda^{*}(\eta)Y_J({\bf x})
\ee
compose the complete set of classical solutions of the Klein--Fock--Gordon
equation~(\ref{varphitilde}) if the Wronskian is equal to
\be
g_\lambda {g_\lambda^{*}}'-{g_\lambda}' g_\lambda^{*}=-2\imath.
\ee

\section{Casimir condensate of a massive scalar field}

The vacuum state is defined by the condition $a_J^{(-)}|0>=0$.
The operator commutation relations are
\be
[a_J^{(-)},a_{J'}^{(+)}]=\delta_{JJ'},\qquad [a_J^{(+)},a_{J'}^{(+)}]=[a_J^{(-)},a_{J'}^{(-)}]=0.
\ee
The quantum field operator is then combined as the sum
\be\label{fieldop}
\tilde\varphi (x)=\sum\limits_J\left(\tilde\varphi_J^{(-)}(x)a_J^{(-)}+\tilde\varphi_J^{(+)}(x)a_J^{(+)}\right).
\ee
The conditions of orthonormality for the functions read
\be
(\tilde\varphi_J^{(\pm)},\tilde\varphi_{J'}^{(\pm)})=\mp\delta_{JJ'},\qquad (\tilde\varphi_J^{(\pm)},\tilde\varphi_{J'}^{(\mp)})=0,
\ee
where the scalar product of functions is defined over a finite volume $V$ as
\be
(f,g)\equiv\int_V\, d^3x\sqrt\gamma\left(f^*\frac{\partial}{\partial t}g
- g^*\frac{\partial}{\partial t}f\right).
\ee

We define the quantum Casimir condensate as the vacuum mean value of two operators
in one point:
\be\label{confcondscalar}
<0|\tilde\varphi (x)\tilde\varphi (x)|0>=
\sum\limits_{J,J'}\tilde\varphi_J^{(-)}(x)\tilde\varphi_{J'}^{(+)}(x)<0|a_J^{(-)}a_{J'}^{(+)}|0>=
\frac{1}{2}\sum\limits_J|g_\lambda(\eta)|^2|Y_J({\bf x})|^2.
\ee
By applying the formula for summation over quantum numbers $(l,M)$
\be\nonumber
\sum\limits_{l,M}|Y_J({\bf x})|^2=\frac{\lambda^2}{2\pi^2},
\ee
we get
\be\label{bosoncond}
<0|\tilde\varphi (x)\tilde\varphi (x)|0>=\frac{1}{4\pi^2}\sum_{\lambda=1}^\infty\lambda^2|g_\lambda (\eta)|^2.
\ee

\begin{figure}
\centerline{\includegraphics[width=4.5in]{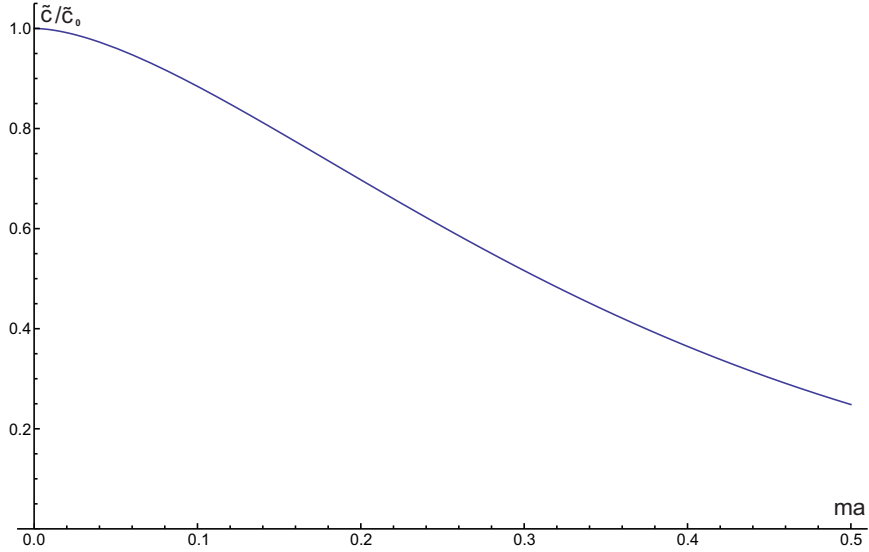}}
\vspace*{8pt}
\caption{Casimir condensate of a conformal massive scalar field
as a function of $ma$. \protect\label{Bosoncondensate}}
\end{figure}

In this paper the effects of vacuum polarization and particle creation are not
considered. Restricting our attention to the pure static Casimir contribution into
the quantum condensate in the case of the quasi-static condition
$(a' = 0)$~\cite{MTrunov}, the solution to the oscillator equation~(\ref{glambda})
can be given as
\be
g_\lambda (\eta)=\frac{e^{\imath\omega_\lambda\eta}}{\sqrt{\omega_\lambda}}.
\ee
Thus one gets the divergent series
\be \label{bosonconda0}
<0|\tilde\varphi (x)\tilde\varphi (x)|0>
=\frac{1}{4\pi^2}\sum_{\lambda=1}^\infty\frac{\lambda^2}{\sqrt{\lambda^2+m^2a^2}}.
\ee
Now the problem is formulated for a scalar field with mass $(ma)$ in a static
Einstein universe $\mathbb{R}^1\times S^3$, where $S^3$ is a sphere of radius $a_0$.
For renormalization of the series the Abel--Plana formula can be applied~\cite{MTrunov}
\be \label{Abel}
{\rm ren}\left\{\sum_{\lambda=0}^\infty F (\lambda)\right\}\equiv
\sum\limits_{\lambda=0}^\infty F(\lambda)-\int\limits_0^\infty dt\, F(t)=\frac{F(0)}{2}+\imath\int\limits_0^\infty dt\,\frac{F(\imath t)-F(-\imath t)}{\exp (2\pi t)-1},
\ee
where $F(\lambda)$ is an analytic function. From the physical point of view,
we have a finite difference between the divergent sum of $F(\lambda)$ defined
in the Friedmann spacetime and the corresponding divergent integral of this
function $F(t)$ defined in the tangent Minkowskian one.
Subtraction of the contribution of the quantum condensate in the
3-dimensional Minkowskian space tangent to the $S^3$, leads to the renormalized
result in the form of the following convergent integral:
\be \label{bosonconda0ren}
\tilde{c}:= <0|\tilde\varphi (x)\tilde\varphi (x)|0>^{\rm ren}
= - \frac{1}{2\pi^2}\int\limits_{ma}^\infty
\frac{\lambda^2 d\lambda}{\sqrt{\lambda^2-m^2a^2}(\exp (2\pi\lambda)-1)}.
\ee
In derivation of this formula, we kept in mind that the integrand function
$F(\lambda)=\lambda^2/\sqrt{\lambda^2+m^2a^2}$ has branch points for
$\lambda=\pm\imath ma$, so
\bea
F(\imath\lambda)&=&F(-\imath\lambda)=-\frac{\lambda^2}{\sqrt{m^2a^2-\lambda^2}},\qquad \lambda<ma,\nonumber\\
F(\imath\lambda)&=&-F(-\imath\lambda)=\imath\frac{\lambda^2}{\sqrt{\lambda^2 -m^2a^2}},\qquad \lambda>ma.\nonumber
\eea
Integral~(\ref{bosonconda0ren}) is limited from the bottom by the variable
bare mass of the elementary particle. For the case of massless virtual particles,
the regularized function in~(\ref{bosonconda0ren}) coincides with
distribution function for bosons with the effective temperature
$T_{\rm eff}=\hbar c/(2\pi k_B)$, where $k_B$ is the Boltzmann constant.
For the massless case $(m=0)$, integration of (\ref{bosonconda0ren}) gives
\be \label{bosoncondmassless}
\tilde{c}_0=-\frac{1}{2\pi^2}\int\limits_{0}^\infty
\frac{\lambda d\lambda}{\exp (2\pi\lambda)-1}=\frac{1}{4\pi^2}{\rm ren}
\sum\limits_{\lambda=1}^\infty \lambda = -\frac{1}{48\pi^2}.
\ee
For elementary particles with $ma \gtrsim 0.5$, the dominant energy condition
$\epsilon\ge |p|$ is violated as discussed in Ref.~\cite{Ford2}. 
This can lead to particle creation by gravitational field.
The dependence of the dimensionless ratio of the conformal
condensates~(\ref{bosonconda0ren}) and (\ref{bosoncondmassless})
on the dimensionless variable mass
${\tilde{c}}/{\tilde{c}_0}= {\tilde{c}}/{\tilde{c}_0}(ma)$ is regular, it
is shown in Fig.~\ref{Bosoncondensate}.

The renormalized value of the static Casimir energy density of the conformal scalar
field~\cite{MTrunov} as the renormalized mean value of the corresponding component
of the conformal energy-momentum tensor reads
\be \label{epsenergy}
\tilde\epsilon := - \tilde{g}^{00}<0|\tilde{T}_{00}|0>^{\rm ren}=
\frac{\hbar c}{2\pi^2a_0^2}\int\limits_{ma}^\infty
d\lambda\,\frac{\lambda^2\sqrt{\lambda^2-m^2a^2}}{\exp (2\pi\lambda)-1}.
\ee
The condensate $\tilde{c}$ (\ref{bosonconda0ren}) and the corresponding energy
density $\tilde\epsilon$ (\ref{epsenergy}) of the massive conformal scalar field
are related to each other. If one differentiates (\ref{epsenergy}) with respect to
the variable mass as a parameter, the relation is revealed
\be \label{dedma}
\frac{\partial\tilde\epsilon}{\partial (ma)}=-(ma)
\frac{1}{2\pi^2a_0^2}\int\limits_{ma}^\infty
\frac{\lambda^2 d\lambda}{\sqrt{\lambda^2-m^2a^2}(\exp (2\pi\lambda)-1)}=\hbar c\left(\frac{ma}{a_0^2}\right)\tilde{c}.
\ee
Formula (\ref{dedma}) can be rewritten also as
\be
\frac{\partial\tilde\epsilon}{\partial (m^2a^2)}=\frac{\hbar c}{2a_0^2}\tilde{c}.
\ee

\section{Conclusions}

In this way, we calculated the static Casimir condensate of a conformal massive scalar
in a compact Friedmann universe. The differential relation between the energy
density and the condensate is obtained.
The Abel--Plana formula, used to renormalize the results, has
a clear physical meaning. It is much more simple than
the regularization procedure with a wavelength cutoff applied in Ref.~\cite{Ford2}.
The applied procedure can be extended for the case of spinor and vector fields.
It would be also interesting to go beyond the conformally static approximation
and consider the dynamics of the Casimir condensate depending on the universe
scale factor $a(\eta)$ evolution.


\end{document}